\documentclass[12pt,a4paper]{article}
\usepackage{amssymb,amsfonts,amsmath}

\usepackage{epsfig, euscript, graphics}

\usepackage[toc,page]{appendix}

\begin{document}

\title{Introductory review to quantum information retrieval}
\author{Alexander Lebedev and Andrei Khrennikov\\ Mathematical Institute,
Linnaeus University, 
 V\"axj\"o, SE-351 95, Sweden}
\date{}

\abstract{Recently people started to understand that applications of the mathematical formalism of quantum theory are not reduced to physics. Nowadays, this formalism is widely used outside of quantum physics, in particular, in cognition, psychology, decision making, information processing, especially information retrieval. The latter is very promising. The aim of this brief introductory review is to stimulate research in this exciting area of information science. This paper is not aimed to present a complete review on the state of art in quantum information retrieval.}

{\bf keywords:} quantum theory; quantum information retrieval; Text access; contexts

\maketitle

\section{Introduction}

During the recent years the mathematical formalism of quantum theory started to be actively applied to a variety of problems outside of physics (see, e.g., \cite{HK}). In particular, applications to information retrieval are very promising, see, e.g.,  book \cite{AE} for the collection of recent papers and articles \cite{AE2}-\cite{AE1}. The aim of this brief introductory  review is to stimulate non-experts in information retrieval to start working in this area of research, especially  those from quantum information theory and those working with quantum-like models in decision making. For these researchers, the paper is endowed with brief introduction to basics of classical information retrieval (see appendixes). On the other hand, quantum methods can be useful for experts in classical information retrieval without knowledge of quantum theory;  for them, we recommend paper \cite{AE3a} providing a short introduction in quantum formalism with emphasis to quantum probability, the main mathematical tool of quantum information retrieval.      

We recall that information retrieval is the activity performed by a computer system of retrieving documents which are relevant to the user's information need. Since the data in the case of information retrieval are unstructured unlike the data of the databases, one can not prove mathematically that one method is better than the other. Therefore, we have to rely on an evaluation that includes users.
We will describe further natural language content analysis, since before process any data further, computers should understand it to a certain extent. 

\subsection{Natural language processing}
It should be noted that not all of the stages of NLP are necessarily applied to a document during the work of IR system. But in general, to understand the structure of the sentence computer does the following steps.
While processing a document at first we do lexical analysis or part-of-speech tagging, assigning to each word its part of speech. After that, we can do Syntactic analysis(Parsing) and it can give us the structure of the sentence. To get the meaning of the sentence it is necessary to do semantic analysis. Moreover, since a lot of information we mean as a matter of course it is difficult to understand for a computer the meaning of a sentence. The main reason for misunderstanding at that level is an ambiguity.
In many cases retrieval systems today use very moderate NLP-techniques, in particular, bag-of-words representation, where all the words are collected together but without preserving word order.

\subsection{Text access}
There exist two different ways of access: Pull mode and Push mode. An example of the former is a search engine, and the example of the latter is a recommender system. The main difference between these is that initiative comes from the user in Pull mode and system takes action in Push mode. Moreover, the information need either supposed to be stable or known to a certain extent in Push mode and to be ad-hoc IN in case of Pull mode.

\subsection{Formal formulation of the Text Retrieval problem}
Since information search in general deals with and is evaluated with the collections of documents that are text, we temporarily focus further on text retrieval.
We have a language dictionary $V =\{w_1, ..., w_N\}$, a query $q = q_1, ..., q_m$, where each word $q_i$ belongs to a dictionary $V$, a collection of documents $C=\{d_1, ..., d_M\}$, where each document $d_i=d_{i1}, ..., d_{im_j}$ is a sequence of words $d_{ij}$ from the dictionary. Besides, there is a subset $R(q)$ in this collection $C$ that are relevant documents.This set $R(q)$ is generally unknown, and in addition, it depends on the specific user. The task of the text retrieval system is to compute an approximation $R'(q)$ of a set of relevant documents.

\subsection{Two strategies for finding R'(q)}
There are two strategies for computing this approximating set. The first strategy is called Document selection, in this case, the binary function is introduced, which is an indicator of the set of relevant documents. The system must decide whether the documents are relevant or not, that is, it is required to know precisely whether the document will be useful to the user or not.

Another strategy, called document ranking, is that there is a real-valued function $f(d, q)$ that determines the extent to which the document is relevant. Accordingly, in this case, the set of approximating $R'=\{d \in C|f(d, q)>\theta\}$ are all documents for which the value of this function is greater than the threshold $\theta$.

The document ranking is generally preferred over document selection as the only knowledge of user's information need is the query. However, the query can be incorrectly formulated  as it can be overly-constrained or under-constrained. In the first case, this will result in that it will not find documents at all,  in the second case, on the contrary, there will be too many documents from which it will be difficult for the user to select the relevant ones.
However, knowledge of all the relevant documents does not solve the difficulties related to the fact that documents can be relevant to different degrees and this problem is solved by ranking. Moreover, there is a theoretical justification for ranking which is called the Probability Ranking Principle\cite{Rob}.

\subsection{Quantum IR}
The quantum mechanical approach to information retrieval began with the pioneering work of Rijsbergen\cite{Rij}, in which he proposed a new point of view on the methods and concepts of information retrieval that already existed at that time, using the formalism of quantum mechanics. In particular, he proposed reformulating the problem of finding a ranking function corresponding to a query into the problem of finding a density operator corresponding to it. Further, his work was continued and supported by Bruza\cite{Bruza}, as well as with working systems proposed by Melucci group\cite{Mela},\cite{Melb}, Piwowarski et al.\cite{Piwo} and developed further by Frommholz et al.\cite{Fromm} This approach offered a new perspective on existing methods of information retrieval with the potential to include all existing methods or to develop new algorithms that would be inspired by a quantum mechanical approach.

We begin with a book with a description and basic ideas of the book Risbergen.
Then we describe methods of HAL and LSI, and proceed further to the concept of Quantum Information Retrieval framework developed by Piwowarski et al., then we proceed the the notion of contexts introduced by Melucci.
Apendices are devoted to the description of classical text retrieval methods.
\section{The geometry of information retrieval}
The author suggests to consider probability for a document to be relevant to the request as a quantum-mechanical observable. More specifically, Rijsbergen assumes that documents corespond to subspaces in a certain enclosing space, more specifically, Hilbert space, which is not necessarily finite-dimensional. Rijsbergen separately emphasizes the potential of applying this theory to quantum information retrieval carried out on a quantum computer. Since the probability of a document being relevant to a given query, in this case, is a probability measure, he suggests the use of the Gleason theorem. Here we cite this theorem verbatim from \cite{hug}

Let $\mu$ be any measure on the closed subspaces of a separable (real or complex) Hilbert space $\mathcal{H}$ of dimension at least $3 .$ There exists a positive self-adjoint operator $\mathbf{T}$ of the trace class such that, for all closed subspaces $L$ of $\mathcal{H}$,
$$
\mu(L)=\operatorname{Tr}\left(\mathbf{T} P_{L}\right)
$$

Gleason's theorem says that every measure defined on closed subspaces can be approximated by a Hermitian operator on these closed spaces.

For each request, in this case, there is its Hermitian operator representing this request. This operator is called the density operator. In addition, it has its orthogonal basis of eigenvectors, while all its eigenvalues are real.
This density operator is the representation of the query; moreover, different bases of eigenvectors correspond to different operators, which, in particular, reflects the subjectivity of relevance. Thus, different bases correspond to different views on the same objects.

Aboutness.
As Rijsbergen emphasizes, it is philosophically unclear whether this is a well-defined concept for information retrieval, because the authors \cite{abruz} cite the fact that there are at least three different definitions of aboutness, nevertheless, for their purposes, Rijsbergen assumes it to be well defined.

In addition, Rijsbergen emphasizes that this concept arises from the need for abstract reasoning about the properties of documents and queries in terms of index terms. But usually, in all cases when we are dealing with a logic-based approach to information retrieval, aboutness is something that is connected with terms.

About the connection between aboutness and relevance, Rijsbergen writes that we would like the corresponding observables to be the same. However, unfortunately, this is not so in reality, since relevance is subjective and depends on the particular user of the system, and aboutness is related to the terms used in the text.

Therefore, for these observables, there is a certain angle between their own bases of eigenvectors.

Therefore, generally speaking, aboutness and relevance do not necessarily commute, however, we assume that observables corresponding to different terms commute.
The former may lead to the situation when the document may appear to be relevant after the aboutness was measured, although it was not relevant on measurement before the measurement of the aboutness.

Thus, in addition to the proposed use of quantum formalism, the main idea is to reformulate the problem from searching for ranking functions (the probability functions of the document to be relevant to the query) to the search of the corresponding density operator mentioned in the Gleason's theorem, and further approaches by \cite{Piwo} are to clarify this approach.

Now we are approaching the concept of semantic spaces that were invented in the information retrieval\cite{sch}. There are various implementations of these semantic spaces; there is a latent semantic analysis \cite{lsa}using a singular value decomposition. Moreover, there is a hypertext analogue of the language\cite{hal}, which in turn requires consideration of the window around each word. However, what the result will be, namely, the matrix of semantic space, very much depends on which method is employed.

Bruza and Cole \cite{Bruza} considered semantic spaces in the context of quantum information retrieval. Moreover, they did experiments with the collection Reuters-21538.

 They examined semantic spaces and considered them to be the sum of unknown semantic spaces corresponding, as they wrote, to each text box of a fixed length that is centered around a given word.
 $$S_{w}=\sum_{j=1}^{k} S_{j}$$

But since there is one obvious thing about such windows, in fact, they most likely had in mind the different meanings and contexts that stood behind it. In their article, they rely on the use of the Gardenfors model of cognition.\cite{gar} According to it, there are three levels of cognition, and the second level is where the geometric structures are located. They used the Gardenfors model of cognition to find the basis for their models and experiments.

Accordingly, they suggested that contextual effects at a conceptual level of cognition can be formalized using the definition of quantum collapse, that is, as a result of measuring the state of the system and provided that we get some specific result, the state of the system is in a pure state, corresponding to its vector corresponding, in turn, to this result that we obtained.

To search for individual semantic spaces, as Bruza and Cole write, they assume that for the semantic space in which a given word is located, this space is a weighted sum of semantic spaces.
$$\begin{aligned} S_{w} &=\sum_{i=1}^{k}\left|e_{i}\right\rangle d_{i}\left\langle e_{i}\right| \\ &=\sum_{i=1}^{k} d_{i}\left|e_{i}\right\rangle\left\langle e_{i}\right| \\ &=d_{1}\left|e_{1}\right\rangle\left\langle e_{1}\left|+\ldots+d_{k}\right| e_{k}\right\rangle\left\langle e_{k}\right| \end{aligned}$$

They made up the density operator for this semantic space, and to find its vectors and eigenvalues, and then considered them and interpret them so that the components of vectors with a positive value and a large magnitude value make sense, but negative ones with a small magnitude value does not make sense in semantic space.

\section{The Quantum information retrieval framework}
The following is a description of the methodology for constructing the space of information needs and the framework of quantum information retrieval proposed by Piwowarski et al. \cite{Piwo} \cite{Piwo1} \cite{Piwo2}
In the approach of Piwowarski et al. there is a resemblance to LSI since spectral analysis is used to obtain the representation of queries and documents.

The key concept is the space of information need(IN), which is considered to be a finite-dimensional Hilbert space. Events, such as document relevance or observed user interactions correspond to subspaces, the density operator is used to represent the current point of view of the information retrieval system on the user's information need.
Moreover, the tensor products of spaces employed to capture various aspects of the informational need.

 Piwowarski et al. wanted to create a user-oriented model and tried to solve the problem of the absence of a unified framework of IR which simultaneously allows solving various challenges that arise in information retrieval.
In particular, they wanted to address such aspects as interaction, diversity, and novelty.

In \cite{Piwo2} information need is presented as a weighted set of vectors that evolves as a user interacts(with the system). The probability of a document to be relevant is taken into account with regard to this set.

 Concerning IN space, mentioned above, each vector in space corresponds to an information need that fully characterizes the possible information need of the user. Each vector of the basis in IN space is called pure information need. Any probabilistic event is represented as a subspace in a Hilbert space.

Probability is first determined for pure information needs, i.e. basis vectors of space.
Traditionally, this probability for a quantum-mechanical model is determined by calculating the squared length of the projection of the vector of pure informational need onto the subspace.
If the event is that the document is relevant, then this probability is the desired ranking function.

 The authors suggest that each vector $\varphi $ of pure information need is associated with a corresponding probability weight
 $\mathcal{V}(\varphi) \in \mathbb{R}$.
Using this formula, we calculate the probability for a set of pure information needs.
$$ 
\operatorname{Pr}(S | \mathcal{V}) =\sum_{\varphi \in \mathcal{V}} \mathcal{V}(\varphi) \operatorname{Pr}(S | \varphi)=\sum_{\varphi \in \mathcal{V}} \mathcal{V}(\varphi)\|\widehat{S} \varphi\|^{2} \\ =\operatorname{tr}\left(\rho_{\mathcal{V}} \widehat{S}\right),
$$ 
where 
$$ 
\rho_{\mathcal{V}}=\sum_{\varphi \in \mathcal{V}} \mathcal{V}(\varphi) \varphi \varphi^{\top}. 
$$

For each document, the authors calculated a projector on it and, for each query, the density operator approximating the corresponding density operator.

In addition, in the experiments the authors considered that the space of pure information needs is a terms' space.

To build a projector onto a document space, Piwowarski et al. used the sliding window approach.
They tried to divide the texts into disjoint pieces of various lengths.
According to the authors, the best approach was shown in which the text was divided into disjoint sentences, however, since it was not always possible to recognize sentences, the authors used the approach of dividing the text into equal parts.
How the authors built the projector on a document?

Each document answers (is relevant) to many requests.
The authors assumed that for each document there is a mapping between the fragments of the document and the set of pure information needs, and the authors hypothesize that the document is relevant to the pure information need if the vector of pure information need is contained in the span of vectors from the set associated with the document.
To present a query from a single term, the authors assumed that the query can be represented as a set $U_t$ of pure information needs vectors corresponding to fragments of the document containing this term.

Since it is not possible to distinguish different vectors from the set $U_t$, the probability of each vector is considered the same, the following density operator corresponds to this term.

$$\rho_{t}=\frac{1}{N_{t}} \sum_{\varphi \in \mathcal{U}_{t}} \varphi \varphi^{\top}$$
The principle is that the more intersection between the vector space and the set of vectors matching the query, the higher the likelihood of relevance.
 For a query consisting of several terms, several approaches were used, the first of which was called a mixture. In this case, the authors suggest that the relevant document should equally meet all the pure information needs associated with each query term.
To calculate the probability of being relevant, one term is first selected from the query, with some probability, and then one of the vectors of the set corresponding to this term.

 Then the process is repeated and the average of all combinations is taken.
The result is the following weighted sum of density operators.
$$
\rho_{q}^{(m)}=\sum_{t \in q} \sum_{\varphi \in \mathcal{U}_{t}} \frac{w_{t}}{N_{t}} \varphi \varphi^{\top}=\sum_{t \in q} w_{t} \rho_{t}
$$
Besides, a different approach was used to present the request, called a mixture of superpositions.
One vector is selected that corresponds to each term with a weight of the root of $w,$ and for this document, we study the question of what is the relevance for a given information need.
For a given information need, accordingly, this process can be repeated for any possible combination of vectors and the final density matrix is equal :
$$
\rho_{q}^{(m s)}=\frac{1}{Z_{q}} \sum_{\varphi_{1} \in \mathcal{U}_{t_{1}}} \cdots \sum_{\varphi_{n} \in \mathcal{U}_{t_{n}}}\left(\sum_{i=1}^{n} \sqrt{\frac{w_{t_{i}}}{N_{t_{i}}}} \varphi_{i}\right)\left(\sum_{i=1}^{n} \sqrt{\frac{w_{t_{i}}}{N_{t_{i}}}} \varphi_{i}\right)^{\top}
$$
Here the weights are parameters that changed during the process and they were selected in some way to improve the result based on tf-idf.

The probability of relevance of a document represented by an information need's vector is the length of the projection of the presentation of the document onto the presentation of information need.
Ingwersen advocated for the provision of information needs as a dynamic component in \cite{turn}.

In addition, the authors considered such a concept as aspects of a pure information need. The authors note that the information need of the user consists of several aspects to which this document should respond.
 Each aspect can be defined in the corresponding space where state vectors are already called pure information need aspects.
An example of such an aspects space is the topical space corresponding to the standard space of terms in which each term corresponds to some dimension. The idea of these authors is to consider the information need as a system consisting of many parts, where each component reflects one aspect of the information need.

 Since each aspect is expressed in its Hilbert space, the entire space of information needs is already a tensor product of Hilbert spaces.
There is also pure IN aspect ``don't care''.
 In cases where some aspect is not important for consideration, the corresponding component is replaced by the ``don't care'' vector, for which the projection onto the subspace corresponding to any document and the corresponding probability is equal to unity.

 
\section{Contexts}

Melucci uses a subspace to represent the context, a concept related to the information need of the user.
Melucci's context approach was proposed in article \cite{Melc}. Various methods for choosing basis vectors in a standard term space can be used to convey context in a vector space model and, in particular, express information needs.
Moreover, the choice of basis for the query vector and the choice of basis for the document vector do not have to coincide. Moreover, these choices can be an expression of the context of the author of the request and expression of the context of the author of the document.
This idea reflects the fact that while these queries can have the same coordinates in different bases of the vector space, one of these queries may be closer to the other in this document in the sense of a scalar product. The following describes the evolution of contexts, how to find the correlation matrix, and so on.

\section{Concluding remarks}

We hope that this short presentation of basics of classical and quantum mathematical modeling of information retrieval can be useful  
for newcomers to this area of research.  As the next step,  for further reading we recommend collection of papers in book \cite{AE}. 

\section*{Acknowledgments} 
This paper was written as a part of the EU-project QUARTZ,  ``Quantum Information Access and Retrieval Theory'' - an Innovative Training Network.

\begin{appendices}

\subsection{Probability Ranking Principle}

Returning the ranked list of documents in the descending order of the probability that the document is relevant to the query is the optimal strategy under the following assumptions: the utility of the document to the user does not depend on the utility of any other document, and the second is that a user reviews the results sequentially. 

It is easy to see that these assumptions are not generally satisfied as, e.g., the utility of a document may depend on the utility of other documents, a user can skip some items.
\section{Overview of text retrieval methods}
The text retrieval model provides us with a computational formalization of relevance. All the classic models below are based on the assumption of using a bag-of-words representation of the text.
\subsection{Similarity-based models}
This kind of models is based on the idea that $f(q, d)=similarity(q, d)$. That is, the underlying assumption that underlies this method is that if the first document and query are more similar to each other than the second document and the query, then the first document is more relevant than the second. An example of such a model is VSM. 
\subsubsection{Vector space model}
A vector model is a framework in which a term vector represents documents or a query. The term is a fundamental concept, for example, a word or phrase. Each term defines one dimension, N terms define an n-dimensional space. The query vector $q=(x_1, ..., x_N)$ consists of coordinates where each coordinate $x_i$is the weight of the term. The vector of the document $d=(y_1, ..., y_N)$ also consists of coordinates, each of which is the weight of the document term.

VSM does not say what the weight of query terms should be, how to place queries in space. The model does not say how to put the document in space, what the weight of the terms of documents should be, and in the end, it does not say what the similarity measure should be between the query and documents.

\paragraph{The simplest instantiation of VSM}
The simplest instantiation of the Vector Space Model implementation is the case where the weights of terms are the indicator function of a term, and we use the usual scalar product of vectors as a measure of similarity.
Well, there are a few problems. First, we do not take into account the fact that the query words are repeated several times in some documents, and most likely those documents are more relevant than those that do not have repetitions. Besides, there are differences in the value of words. Some words such as articles or prepositions are apparently of less value than nouns adjectives. 
Hence we need to modify the most straightforward implementation in order to improve the performance.

\paragraph{Parameters to count}

With the assumption of using bag-of-words, if we consider a query consisting of several words and a given document, then we can consider the functions of each of the words in the query and the document. We can try to characterize the popularity of the term in the collection.
The parameters that we can extract are the following, for example, term frequency(TF), i.e., how many times we meet the word from the query(or the term in case of information retrieval) in the document, the document length $|d|$, document frequency $df(w)$--- how often does the word $w$ appear in the entire collection c.

\paragraph{Improved VSM}
In order to solve the problem related to the fact that some query words are repeated in the document, we multiply the term weight in the document vector by the frequency of appearance of this term in the document(TF). In order to cope with the problem of frequent words, such as prepositions and articles, we can use Inverse Document Frequency(IDF), by doing this we will award the terms that are rarely found in the collection of documents.
IDF can be defined as 
$$IDF(W) = log(\frac{M+1}{k}),$$
where k is total number of documents containing word W(document frequency), $M$ is the number of documents in the collection.
\paragraph{Ranking with TF-IDF weighting}
In this improved VSM we will have the following ranking function:
$$f(q, d)=\sum_{i=1}^{N}x_i y_i= \sum_{w \in q \cap d}c(w, q)c(w, d)\log \frac{M + 1}{df(w)},$$ where $c(w, *)$ is a counter of the number of occurrence of word w in q or d.
\paragraph{TF Transformation}
There are many transformations that are applicable to the function of the counter. In particular, we can use a function that is equal to one for a counter greater than one, then we'll just get the simplest VSM model. Further, there exists an optimal transformation, called BM25(although the formula were deduced from probabilistic model), and the formula is as follows:
$$\frac{(k+1)x}{x+k},$$
where x is a counter c(w, d), and k is the parameter. By changing the parameter, we can obtain the extremal values of this function: piecewise constant and for large parameter values the straight line $y = x$.
\paragraph{Document length normalization}  Length normalization allows us to take into account the length of the documents. We take the average length of the document in the collection and penalize documents that are longer and reward those that are shorter.
$$normalizer = 1 - b + b\frac{|d|}{avdl}$$

These ideas lead us to ranking functions of best performance:
\paragraph{Pivoted length normalization}
Singhal et al. \cite{Sin} proposed the following the state of art ranking function:
$$f(q, d) = \sum_{w\in q \cap d}c(w, q)\frac{\ln[1 + \ln[1+c(w, d)]] }{1 - b + b\frac{|d|}{avdl}}\log \frac{M + 1}{df(w)},$$
\paragraph{BM25}
Robertson \& Walker \cite{bm}:
$$f(q, d) = \sum_{w\in q \cap d}c(w, q)\frac{(k+1)c(w, q)}{c(w, q)+k(1 - b + b\frac{|d|}{avdl})}\log \frac{M + 1}{df(w)}$$
\paragraph{Further development}
We can improve the VSM further. To clarify the concept of dimension and the concept of the similarity function, in particular, to refine the notion of measurement such NLP methods as root extraction, stopwords removal can be used, stemming.   In addition, phrases can be used as a term. As a similarity function, the other scalar product can be chosen or one can also choose a similarity function that is not associated with a scalar product. However, practice shows that the scalar product gives so far the best results.

BM25 can be improved in the sense of working with particular kinds of data, for example, documents with structures, with fields.
\subsection{Probabilistic models}
With this approach, both documents and requests are considered as observations of random variables. In this method $f(q, d) = p(R=1|d, q)$, where $R$ is binary random variable, $p(R=1|d, q)$ is called score of a document with respect to a query. There are a few models which use this idea:

\begin{itemize}
	\item Classic probabilistic model
	\item Language model
	\item Divergence-from-randomness model
	\item Probabilistic inference model
\end{itemize}
\subsection{Axiomatic model} The ranking function $f(q, d)$ must satisfy a set of axioms.

\section{Model of best performance}
According to Fang et al. \cite{fan} models BM25, Pivoted length normalization, Query likelihood, PL2 tend to perform equally well.
\section{Implementation of text retrival system}
Usually, the information retrieval system consists of 3 parts. In the first part, the documents are indexed, that is,  a special data structure called an index is constructed, and this data-structure helps to quickly respond to user requests. In the second part, scoring takes place, and in the third part feedback is taken into account.
\subsection{Preprocessing}
Preprocessing of the document includes tokenization, that is, the division of sentences into, for example, words, but taking into account language specific features. At the same time, words with the same meaning are mapped in the same word. Further, it followed by stemming, i.e., mapping words into the root.
In order to avoid a situation when the corresponding  lists are excessively long, stopwords (very frequent words such as articles etc) are also removed before building the index.
After the preprocessing is done the index can be built. Often, an inverted index is used in which for any term there exist a list of document id's in which this term occurs.
\subsection{Storage}
For large collections, the use of posting lists requires storage on disk and not in RAM. Therefore, optimization is required not only to save disk space but because the recording and input processes require much more time than the processes inside the RAM.
Unfortunately, in most applications, the size of the dataset is large and therefore it is impossible to store posting lists in RAM.
To reduce the amount of memory needed to store an index, several methods are used. For example, we store not the id of the documents, but the difference between the id, which is a small number. In addition, various encodings could be used such as unitary, gamma, delta, and many others.
\subsection{Sorting inverted index}
A known method of creating an inverted index is a sorting method. For each term in the document, we create a tuple, which includes the index of the term (we consider the terms numbered) and the document id and the number of times this term has been encountered. Then we divide the set of all thus created tuples into sets that fit in the RAM, sort these files already by the term index. We create many such files and then merge such files among themselves.
\section{Evaluation of IR systems}
To compare various information retrieval systems, as well as to understand how useful information retrieval systems are in general, we need to be able to evaluate IR system.
\subsection{The Cranfield evaluation methodology}
The main idea of this methodology is to build a reusable test collection of documents and queries as well as judgments of relevance, ideally created by the same people who formulated queries.
Having such a collection of documents and queries, in order to evaluate the system it is necessary to use different metrics since the tasks facing the user can vary.
\subsection{Metrics}
The precision metric shows the ratio of the number of relevant documents to the total number of documents in the search results. Recall metric shows how many of the relevant documents appeared in the search results. Different measures are built employing precision and recall, for example, the F-measure, which is a parameterized harmonic average of precision and recall. The harmonic mean is used so that there is no situation in which one parameter is small and the other is large, and the total metric is large enough.
$$F_{\beta}= \frac{1}{\frac{\beta^2}{1 + \beta^2}\frac{1}{R}+\frac{1}{1+\beta^2}\frac{1}{P}}$$
\end{appendices}

\end{document}